\documentclass[aps,prb,superscriptaddress,twocolumn,showpacs]{revtex4}
\usepackage{amsmath}
\usepackage{amssymb}
\usepackage{amsthm}
\usepackage{graphicx}
\usepackage{graphics}
\usepackage{subfigure}
\usepackage{bm}
\usepackage{hyperref}
\usepackage{rotating}
\usepackage[usenames]{color}

\usepackage{epstopdf}
\usepackage{epsf}

\newcommand{\abs}[1]{\left\vert#1\right\vert}

\def\beas{\begin{eqnarray*}}
\def\eeas{\end{eqnarray*}}
\def\bea{\begin{eqnarray}}
\def\eea{\end{eqnarray}}
\def\be{\begin{equation}}
\def\ee{\end{equation}}

\newcommand{\ka}{\kappa}

\begin{document}

\title{Theory of the Fabry-Perot quantum Hall interferometer}%

\author{Bertrand I. Halperin}
\affiliation{Physics Department, Harvard University, Cambridge MA 02138, USA}

\author{Ady Stern}
\affiliation{Department of Condensed Matter Physics, Weizmann
Institute of Science, Rehovot 76100, Israel}

\author{Izhar Neder}
\affiliation{Physics Department, Harvard University, Cambridge MA 02138, USA}

\author{Bernd Rosenow}
\affiliation{Institute for Theoretical Physics, Leipzig University, D-04009 Leipzig, Germany}

\date{\today}
\begin{abstract}
We analyze interference phenomena in the quantum-Hall analog of the Fabry-Perot interferometer, exploring the roles of the Aharonov-Bohm effect,  Coulomb interactions, and  fractional statistics on the  oscillations of the resistance as one varies the magnetic field  $B$ and/or the  voltage  $V_G$ applied to a side gate. Coulomb interactions couple the interfering edge mode to localized quasiparticle  states in the bulk, whose occupation is quantized in integer values.    For the integer quantum Hall effect, if the bulk-edge coupling is absent, the resistance exhibits an  Aharonov-Bohm (AB) periodicity, where the phase is equal to  the number of quanta of magnetic flux enclosed by a specified interferometer area. When bulk-edge coupling is present, the actual area of the interferometer oscillates as function of $B$ and $V_G$, with a combination  of a smooth variation and  abrupt jumps due to  changes in the number of  quasi-particles in  the bulk of the interferometer. This modulates the Aharonov-Bohm phase and gives rise to additional periodicities in the resistance. In the limit of strong interactions,  the amplitude of the AB oscillations becomes negligible, and  one sees only the new ``Coulomb-dominated'' (CD) periodicity.  In the limits where either the AB or the CD periodicities dominate, a color map of resistance  will show a series of parallel stripes in the $B-V_G$ plane, but the two cases show different stripe spacings and slopes of opposite signs. At intermediate coupling, one sees a superposition of the two patterns. We discuss dependences of the interference intensities on parameters including the temperature and the backscattering strengths of the individual constrictions. We also discuss how results are modified in a fractional quantized Hall system, and the extent to which the interferometer may demonstrate the fractional statistics of the quasiparticles.

\end{abstract}

\pacs{73.43.Cd, 73.43.Jn, 85.35.Ds, 73.23.Hk}
\maketitle

\section{Introduction} \label{Introduction}
\subsection{Background}
In the last few years there has been  a surge of interest in electronic interference phenomena in the regime of the quantum Hall effect. This interest, both   theoretical \cite{Chamon+97,Fradkin+98,SteHa06,Bonderson+06a,Bonderson+06b,RoHa07,Ilan+09} and experimental \cite{machzender,Camino+05,Camino+07a,Camino+07b,Godfrey+07,Zhang+09,McClure+09,Ping+09,Ofek+09,Willet+09a,Willet+09b,Alphenaar+92,RD,Ilani+04,Simmons89,Hackens+10}, results in large part from the hope of utilizing interference to probe unconventional statistics in various fractional quantum Hall states.
Interestingly, interferometer experiments have led to puzzling results even in the
integer regime, which have posed a challenge to our theoretical understanding.

Arguably the simplest realization of a quantum Hall interferometer is an analog to the optical Fabry-Perot device. It is constructed of a Hall bar perturbed by two constrictions, each of which introduces an amplitude for inter-edge scattering. (See Figure 1)  The backscattering probability of 
a wave packet  that goes through the constrictions is then determined by an interference of trajectories. In the limit of weak inter-edge scattering, two trajectories interfere, corresponding to scattering across each of the two constrictions. As the scattering amplitudes get larger, multiple reflections play a more significant role.

\begin{figure}[t]
\begin{center}
\includegraphics[width=1\columnwidth]{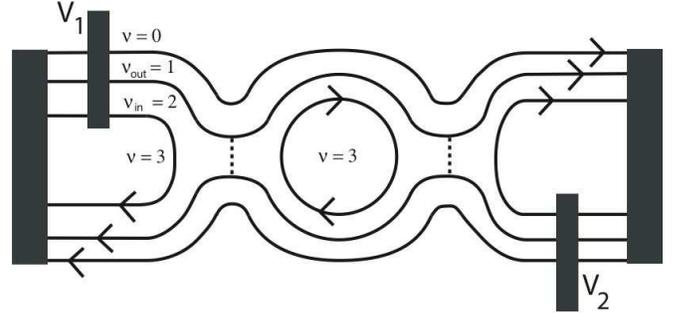}
\label{setup}
\caption{Fabry-Perot interferometer with $f_T =1$ totally transmitted edge modes.  Filling factor at the center of the constriction is in the range $1<\nu_c<2$ and the partially transmitted edge mode separates quantized Hall regions with nominal filling $\nu_{out} = 1$ and $\nu_{\rm{in}} = 2$.  A third edge mode is totally reflected before entering the constrictions, as the bulk filling factor in the center of the interferometer lies in the range $2.5 < \nu_b < 3.5$  Dotted lines in each constriction show the locations of backscattering between the two edges.}
\end{center}
\end{figure}

In our analysis, we assume that the two constrictions forming the interferometer are identical to each other, and that there is a single partially-transmitted edge channel penetrating the two constrictions. This partially transmitted channel separates two quantized Hall states corresponding to rational  filling factors $\nu_{in}> \nu_{out}$, with $ \nu_{out}$ being closer to the sample edge.
In addition to the interfering channel, there may be a number of outer edge channels that are fully transmitted through the two constrictions, whose number we denote by $f_T \geq 0$. The situations considered in this paper assume that the states $\nu_{in}$ and $\nu_{out}$ are either integer states or integers plus a fraction described in the composite fermion picture, where the partially filled Landau level is less than half full. In particular, this means that all edge states propagate in the same direction. We refer to the cases where $\nu_{in} $ is integer or fractional as IQHE and FQHE interferometer, respectively.

For non-interacting electrons, there will be an  interference between electrons backscattered at the two constrictions, with a  relative phase determined by the Aharonov-Bohm effect. It is periodic in the magnetic flux $\Phi$ enclosed by the loop defined by the two interfering trajectories, with a period of one flux quantum $\Phi_0$ (we define the flux quantum as $\Phi_0 = h/|e| > 0$, where $e<0$ is the electron charge.).  For a uniform magnetic field $B$ the flux is $\Phi=BA_I$, with $A_I$ being the area of the interference loop. Experimentally, it is customary to affect this flux through two experimental knobs: $B$ the magnetic field, and $V_G$, the voltage on a gate that affects the area of the loop. The gate may be positioned above the interference loop or to its side. For fractional quantum Hall states, where electron-electron interaction is essential, the relative phase is made of two contributions, an Aharonov-Bohm phase that is scaled down by the charge of the interfering quasi-particle, and an anyonic phase, accumulated when one quasi-particle encircles another.

Experimentally, several remarkable observations were made
\cite{Camino+07a,Godfrey+07,Zhang+09,Ofek+09,Alphenaar+92}   when interference was measured in small Fabry-Perot interferometers, e.g.  with an interference loop whose area is around $5\mu m^2$. One observation was that  when the magnetic field is varied, the
backscattering current oscillates as a function of the magnetic field, but  the period $\Delta B$ of the oscillations was not $\Phi_0/A_I$. Rather, it was given by   $\Phi_0/ f_T A_I$, which means, in particular, that there was  no dependence on $B$ for $f_T=0$.  The period $\Delta B$ did not change when $\nu_c$, the filling factor at the center of the constriction was varied in the range between $f_T$ and $f_T+1$, and the back scattering probability for the partially transmitted edge state varied from strong to weak.
Second, when the lines of constant phase in the $B-V_G$ plane were examined \cite{Zhang+09,Ofek+09}, they were found to have positive slope, which is opposite sign relative to what one would naively expect for an Aharonov-Bohm interference effect (Similar lines were observed also in Ref. [\onlinecite{Ilani+04}], where a scanning probe was used to probe the spectrum of excitations of a spontaneously formed quantum dot).  By contrast, in interferometers that are sufficiently large (e.g., area around $17\mu m^2$), where the center island is covered by a screening top gate, the conventional Aharonov-Bohm pattern was observed, with field period $\Phi_0/A_I$ and negative slope for the lines of constant phase. A similar Aharonov-Bohm behavior was also observed in some small interferometers\cite{Willet+09a,Willet+09b}.

Previous works have explained that the periodicities and slopes in the Fabry-Perot interferometer are affected by the Coulomb interactions and the discreteness of electronic charges \cite{RoHa07,Zhang+09,Ofek+09}. The regime of parameters where lines of constant phase have positive slope (or zero slope in the case $f_T=0$) will be  referred to as the Coulomb-Dominated (CD) regime, in contrast with the Aharonov-Bohm (AB) regime.

In this article, we present a general picture of the interplay of the AB and CD regimes in the Fabry-Perot interferometer, and elucidate the way this interplay is determined by the combination of Coulomb interaction and charge discreteness.  We limit our analysis of the FQHE to abelian states. We hope to
extend our present study to the case of non-abelian states in a future
publication.

\subsection{Summary of our results}

Before we turn into a detailed discussion, we summarize our results and present a physical way of understanding them. Generally, when electron-electron interactions are taken into account, we find that the area $A_I$ enclosed by the interfering edge state is not a smooth monotonic function of the magnetic field and gate voltage.  Rather, we find that $A_I$ has the form
\be
A_I = \bar{A}(B, V_G) +\delta A_I \, ,
\ee
where $\bar{A}$ is a slowly varying function of its arguments, while $\delta A_I$ has rapid oscillations, on the scale of one flux quantum or on a scale of a change in $V_G$ that adds one electron.  (We assume that the area $A_I$ is large enough to enclose  many electrons and  flux quanta, so that the  oscillations occur on a scale where there is only a small fractional change in $B$ or  $\bar{A}$.  We shall also assume, unless otherwise stated, that the secular  area $\bar{A}$  is only weakly dependent on the magnetic field $B$, i.e., that  $ B \partial \bar{A} /\partial B$ is negligible compared to $\bar {A}$.) The oscillatory dependence of $\delta A_I$ on the magnetic field and $V_G$ can have striking consequences on the interference pattern, as we shall see below.

Typically, experiments measure   the ``diagonal resistance" $R_D$ [\onlinecite{RD}],  which is essentially the two-terminal Hall resistance of the interferometer region.
We find that $R_D$ has an oscillatory part $\delta R$, which is a periodic function of $B$ and $\delta V_G$. In the limit of weak backscattering it may be written as
\begin{equation}
\delta R={\rm Re}\ \left( \sum_{m=-\infty}^\infty R_me^{2\pi i (m\phi+\alpha_m\delta V_G) } \right) \, ,
\label{decomposition}
\end{equation}
where
\begin{equation}
\phi \equiv B \bar{A} / \Phi_0 \, ,
\end{equation}
and the coefficients $R_m,\, \alpha_m$ are real and only slowly varying functions of $B, \,V_G$. The voltage $V_G$ affects the phases $e^{2\pi i(m\phi+\alpha_m\delta V_G)}$ in (\ref{decomposition}) in two ways. First, it affects the flux $\phi$ through its effect on the area $\bar A$. Second, it affects the density in the bulk of the interferometer, indirectly affecting the interference through interactions of the edge with the bulk. The coefficients $\alpha_m$ quantify the latter effect, which we will analyze further below.

\begin{figure*}[t]
\centering
\includegraphics*[width=7.0in]{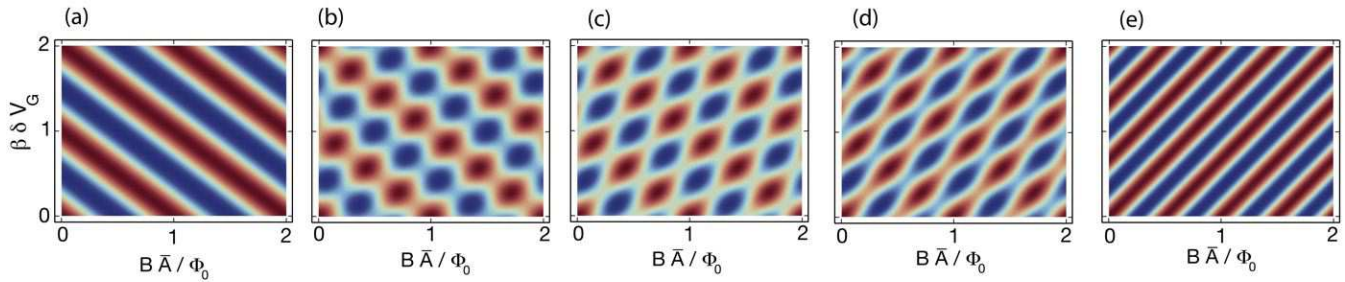}
\caption{$<\delta R>  = {\rm Re} \left(R_1 e^{2 \pi i \phi} + R_{-f_T} e^{-2\pi i f_T \phi}\right)$      as a color map in the plane of $B$ and $V_G$, for  $f_T=2$, with the parameter  $\gamma $ chosen equal to $3.5  \beta$.
Panels (a), (b), (c), (d), and (e) , have, respectively  $|R_{-2}/R_1| =$0, 0.5, 1, 2, and $\infty$ , corresponding to to the AB, mixed and CD regimes.  All
Fourier components other than $m=1$ and $m=-f_T$ are neglected. Alternating red and blue regions represent positive and negative values, respectively, while white signifies a value close to zero. }
\label{result1}
\end{figure*}
%

For non-interacting electrons (the extreme AB limit), the weak backscattering limit has only one non-zero component in Eq. (\ref{decomposition}). That component is $m=1$, with $\alpha_1 =0.$  Since the area $\bar{A}$ should be a monotonically increasing function of $V_G$, we find that for small changes in $B$ and $V_G$
the contours of  of constant phase are  straight lines of negative slope in the $B- V_G$ plane. (When backscattering becomes stronger,  multiple reflections lead to more harmonics of $m$ showing up, but  still $\alpha_m=0$, so the slope does not change.)  When plotted as a color-scale map in a $B-V_G$ plane, the resistance $R_D$ forms a set of parallel lines, such as the dominant features seen in Fig. [2a].

Electron-electron interactions lead to two important differences between the quantum Hall interferometer and a naive Aharonov-Bohm interference experiment. First, as mentioned above, the area $A_I$ of the interference loop is not rigidly constrained a priori, but can fluctuate slightly. Thus, the area of the interference loop varies with magnetic field and the flux within the loop is generally not a simple linear function of the magnetic field. The position of the edge is related to the charge it encloses, and its variation in our model is a consequence of considerations of energy. Second, we model the region enclosed by the interference loop as one in which there are localized states close to the chemical potential. The number $N_L$ of  electrons (in the IQHE regime) or quasi-particles (in the FQHE regime) that are localized in the bulk is an integer, and varies discretely. Due to considerations of energy, an abrupt change of occupation of a localized state as the magnetic field is varied affects also the position of the interfering edge, and hence induces an abrupt change in the flux enclosed by the interference loop.

Thus, as $B$ or $V_G$ vary, the phase accumulated by the interfering particle, $\theta$, evolves in two ways: continuous evolution for as long as $N_L$ does not vary, and abrupt jumps for magnetic fields at which $N_L$ abruptly changes. The continuous change results from the variation of the magnetic flux in the interference loop, both directly as a consequence of the varying $B$, and indirectly as a consequence of the variation of the loop's area $A_I$. The abrupt change results from the effect of a variation of $N_L$ on the area of the interference loop, and, in the FQHE,  from the anyonic phase accumulated when fractionally charged quasi-particles encircle one another. Specifically, in the integer case, $\theta$ is simply related to the field $B$ and the area $A_I$ by
\be
\label{thetaAI}
\theta =  2 \pi B A_{I}/\Phi_0 \, ,
\ee
while, for the FQHE states that we consider, we have
\begin{equation}
\theta=2\pi e^*_{in}\frac{B A_{I}}{\Phi_0}+N_L\theta_a \, ,
\label{fractheta}
\end{equation}
where $\theta_a$ is the phase accumulated when one elementary quasi-particle of charge of the inner  FQHM state $\nu_{in}$  encircles another, and $e^*_{in}$ is the charge of the quasiparticle. Here, and in the following, charge is to be measured in units of the (negative) electron charge $e$.

Within our model, both the rate of continuous evolution of the phase, $d\theta/d\phi$, and the size 
$2 \pi \Delta$ of the phase jump associated with a change of $N_L$ by -1, vary only slowly with $B$ and $V_G$. The same holds for the magnetic field spacings between consecutive changes in $N_L$.

In the extreme Coulomb dominated regime, for integer and fractional states alike, we find that a change of $N_L$ is accompanied by a change of the area of the interference loop in such a way that the phase jump $\Delta\theta$ is an unobservable integer multiple of $2\pi$. Coulomb interaction makes the area vary in such a way that the continuous variation of the phase follows $\frac{d\theta}{d\phi}=- 2 \pi \frac{\nu_{out}}{e^*_{out}}$, where $e^*_{out}$ is the elementary charge of the outer $\nu_{out}$ quantized Hall state. Neglecting the unobservable phase jumps, then, $\theta=- 2 \pi \frac{\nu_{\rm out}}{e^*_{\rm out}}\phi$ for both the IQHE and the FQHE. This limit characterizes interferometers where the capacitive coupling of the bulk and the edge is strong. By contrast, in the extreme Aharonov-Bohm case, where the bulk and the interfering edge are not coupled,  the area of the interference loop does not vary with $B$ at all.   Moreover,  $A_I$ does not vary when $N_L$ varies. Thus, for integer states $\theta=2 \pi \phi$. The fractional case is more complicated due to the anyonic phase $\theta_a$.

In between these two extremes, $\theta$ is not proportional to $\phi$, and thus the Fourier transform of $e^{i\theta}$ with respect to $\phi$ has more than one component. For fractional states this is the case even in the extreme AB limit, due to the anyonic phase $\theta_a$. We find that for all the cases we consider, the components that appear in Eq. (\ref{decomposition}) satisfy
\begin{equation}
m=-\frac{\nu_{out}}{e^*_{out}}+g\frac{\nu_{in}}{e^*_{in}} \, ,
\label{allowedms}
\end{equation}
where  $g$ is an integer.  Note that the ratios $\nu_{out} /e^*_{out}$ and $\nu_{in} / e^*_{in}$ are always integers, so the allowed values of $m$ are integers as well. Moreover, due to the interaction, $\alpha_m$ is  
not proportional to $m$, leading to different slopes of the equal phase lines for the different $m$ components.

The Coulomb dominated limit and the Aharonov-Bohm limit are both defined in terms of the dominant values of $g$ in (\ref{allowedms}). In the extreme CD limit the only term that appears in the sum (\ref{decomposition}) is that of $g=0$ in (\ref{allowedms}), both for integer and fractional states. In the extreme AB limit of integer states the only term that appears in (\ref{decomposition}) is the naive Aharonov-Bohm term $m=1$ (or $g=1$ in (\ref{allowedms})). For fractional states, however, there will be coupling due to the phase jumps associated with the anyonic statistics of the quasi-particles, and one would not find pure AB behavior, (only $g=1$), even when the Coulomb coupling between $N_L$ and $A_I$ can be neglected.  Moreover, for FQHE states with $\nu_{in} > 1$, one finds that there is no value of $g$ that generates $m=1$ in Eq. (\ref{allowedms}), so the naive AB period is completely absent in the weak backscattering limit.

In between the extreme Aharonov-Bohm and Coulomb-dominated limits, all integers $g$ appear in the Fourier decomposition of $\delta R$, with the relative dominance of the AB and CD components being determined by the value of $\Delta$. We find, under plausible assumptions, that $0<\Delta<1$, and that if   $0 \leq \Delta  < 1/2$   the AB term will dominate, whereas the CD term will dominate if $ 1/2< \Delta  \leq 1. $

When the sum (\ref{decomposition}) is dominated by one term, as is the case in the CD limit and the AB limit of the IQHE, the color-scale plot of $\delta R$ on the $B-V_G$ plane is characterized by a set of parallel lines, as is the case in Figs. [2a] and [2e]. 

The three figures [2b]-[2d] show the intermediate case, in which several values of $m$ contribute, and $\alpha_m$ is not proportional to $m$.
Then the structure of $R_D$ in the $B-V_G$ plane assumes a form of a two-dimensional lattice, rather than a set of lines, as it would if  $\alpha_m$ stayed proportional to $m$.
The periodic structure may be characterized by a unit cell in the $B-V_G$ plane, described by two elementary lattice vectors $\bf{b}$ and $\bf{v}$. In the most general case these vectors can have two arbitrary orientations in the plane. However, if the secular  area $\bar{A}$  is only weakly dependent on the magnetic field $B$, that is if $ B \partial \bar{A} /\partial B\ll \bar{A}$, we find that one of the elementary lattice vectors  will be parallel to the $B$ axis.  Specifically,  if $V_G$ is held constant,  $\delta R$ will be unchanged when $B$ is changed by the amount that increases $\phi$ by one. (We emphasize that this is true even if the interfering particles are fractionally charged.) In our later discussions, rather than employing the direct lattice vectors $\bf{b}$ and $\bf{v}$, we shall use a description in terms of their reciprocal lattice vectors.

The restriction of the Fourier harmonics to the values (\ref{allowedms}) is valid only in the limit of weak backscattering. As the constrictions are further closed and the amplitude for backscattering becomes appreciable, all values of $m$ appear in (\ref{decomposition}). In the limit where this amplitude is strong, oscillations in the reflection probability turn into transmission resonances. The spacing between these resonances varies with the degree of coupling between the bulk and the edge. Generally, a transmission resonance occurs when the almost closed interfering edge has a degeneracy point, at which it may accommodate an extra electron (for the IQHE) or quasi-particle (for the FQHE) at no extra energy cost. In the Aharonov-Bohm limit, it is the energy of the edge, decoupled from the localized charges it encloses, that should be invariant to adding an extra charge carrier. At the integer quantum Hall regime, that would give rise to one transmission resonance per every flux quantum. In the Coulomb Dominated limit, when the introduction of localized charges affects the energy of the edge through their mutual coupling, there would be $\nu_{out}/e^*_{out}$ resonances per quantum of flux, in both the IQHE and the FQHE. Thus, the distinction between the AB and CD limits holds even in the limit of a closed interferometer, where the interfering edge almost becomes a quantum dot.

As should be clear from the discussion above, the form of $\delta R$ depends crucially on the continuous and abrupt phase variations $d\theta/d\phi$ and $\Delta\theta$. Both of these quantities depend on energy considerations, since the interferometer's area is a property of thermal equilibrium. We model the energy of the interferometer in terms of a capacitor network. The parameters of the model, describing the self capacitance of the interfering edge, the self capacitance of the localized quasi-particles, the mutual capacitance of the two, and the capacitive coupling of the gate to the interferometer, depend on microscopic parameters which we cannot accurately calculate at this point. However, we are able to give some insights into the way in which various parameters should vary with details of the systems, including particularly the perimeter and area of the interference loop.

\begin{table}
\begin{tabular}{|p{1.4cm}|p{5.8cm}|p{1cm}|}
\hline
symbol & short description &  section \\
\hline
$\nu_{\rm in}$ ($\nu_{\rm out}$) & filling factor inside (outside) the
interfering edge state & I A \\
$f_T$ &  number of fully transmitted edge states & I A \\
$B$ & magnetic field & I A \\
$\Delta B$ & magnetic field periodicity & I A \\
$V_G$ & voltage applied to a gate & I A \\
$A_I$ &  area of the interference loop & I A \\
$\bar A$ &  slowly varying part of $A_I$ & I B \\
$\delta A_I$ &  rapidly oscillating part of $A_I$ & I B \\
$R_D$ & diagonal resistance & I B \\
$\delta R$ &  oscillatory part of $R_D$ & I B \\
$\phi$ & magnetic flux within the area $\bar A$ & I B \\
$\alpha_m$ & quantifies the effect of $V_G$ on the bulk of the interferometer
loop & I B \\
$N_L$ & number of electrons or quasi-particles localized in the bulk of the
interferometer & I B \\
$\theta$ & the interference phase & I B \\
$e^*_{in}$ ($e^*_{\rm out}$) & the charge of a quasi-particle in the $\nu_{in}$
($\nu_{\rm out}$) state & I B \\
$2 \pi \Delta$ & jump in phase $\theta$ when $N_L$ varies by $-1$ & I B \\
$\theta_a$ & anyonic phase & I B \\
\hline
$r_1,r_2$ & reflection amplitudes at constrictions 1,2 & II \\
$N_j^e,N_j^h$ & integer number of localized electrons and holes in the
$j$'th Landau level & II \\
$K_I$, $K_{IL}$, $K_L$ & coupling constants in the energy functional
describing the interferometer & II A \\
$\bar q$ & effective bulk background charge & II A \\
$\beta$ & quantifies the effect of $V_G$ on the area of the interferometer &
II A \\
$\gamma$ & quantifies the effect of $V_G$ on the bulk background charge & II
A \\
$\Delta\nu$ & $\nu_{\rm in}-\nu_{\rm out}$ & II B \\
$C_I$,$C_L$, $C_{IL}$ & re-parametrization of $K_I,K_L, K_{IL}$ by effective capacitances & II C \\
$\mu_I,\mu_L$ & electro-chemical potentials of the $I$ and $L$ regions & II
C \\
$w,L$ & width and length of the region of non-uniform density near the
loop's edge & II C \\
\hline
$Z$ & partition function & III B \\
${\vec G}_{gh}$ & reciprocal lattice vectors of the 2D description of
$\delta R(B,V_G)$ & III E \\
$\lambda $ & describes the variation of $\bar A$ with $B$ & III E \\
$\eta$ & describes the variation of $\bar q$ with $B$ & III E \\
\hline
$P_R$ & reflection probability for the interfering edge state & IV \\
$\chi_\pm$ &interferometer scattering phase shifts & IV \\
$\rho(\epsilon)$ & density of states & IV \\
\hline
$\Delta\phi$ & flux spacing between resonances & V \\
$N_o$ & total number of electrons in the highest Landau level enclosed by
the interfering edge channels  & V \\
\hline
\end{tabular}
\vskip 2mm \caption{List of symbols, their brief description, and the
section where they are defined}
\label{tableofsymbols}
\end{table}

\subsection{The structure of the paper}
The structure of the paper is as follows.  In Sec. (\ref{sec:weak-model}) we deal with the weak backscattering limit. We identify what we believe to be the important degrees of freedom in the interferometer, express the phase $\theta$ in terms of these degrees of freedom and introduce an energy functional in terms of these degrees of freedom. In Sec. (\ref{weak-calc}) we calculate the thermal average of $e^{i\theta}$, which is the factor that determines the interference contribution to $R_D$ in the weak backscattering limit, and distinguish between the Aharonov-Bohm and Coulomb-Dominated limits. In Sec (\ref{intermediatebackscattering}) we extend the discussion to the regime of intermediate backscattering, and in Sec. (\ref{strongbackscattering}) to the regime of strong backscattering. In Sec. (\ref{model}) we exemplify the way in which the energy parameters for the interfering edge and the localized states  can be influenced by coupling to edge states that are fully transmitted, by solving in detail two simple  models. In Sec. (\ref{Literature}) we compare our findings to earlier experimental and theoretical works. Finally, we summarize our results in  Sec. (\ref{summary}).

For the convenience of the reader, we include a table
with a list of the main symbols used in the
paper, their brief description,  and a pointer to the Section in which they
are defined.

\section{The physical model - weak backscattering case\label{sec:weak-model}}

In this section we introduce the physical model on which we base our analysis of the weak backscattering limit. We start with the IQHE interferometers, and then generalize to the FQHE ones. 

In the weak backscattering limit there should be an oscillatory part of the backscattered resistance
given by
\be
\label{intphase}
\delta R \propto {\rm{Re}} [ r_1 r_2^* < e^{i \theta} > ]
\ee
where  $r_1, r_2$ are the reflection amplitudes at  the two  constrictions,  and the angular brackets represent an average over thermal fluctuations.  We focus here on measurements in the limit of small source-drain bias, so we may consider all leads to be at the same electrochemical potential $\mu$.
We assume the change in $B$ and $V_G$ to be small enough that we may neglect any changes in $r_1$ and $r_2$, and associate oscillations in the resistance with oscillations in the phase factor $<e^{i \theta}>$.

Our analysis of the phase factor $e^{i\theta}$ is based on the following picture of the edge of a quantum Hall fluid in the integer regime.
We expect that any Landau level $j$ which is more than half filled in the bulk of the system will have a single chiral edge state that circulates along the edge of the system.  To the extent that the electron density varies smoothly near the edge of the sample, on the scale of the magnetic length, we expect that  the spatial location of the edge state will be close to the point where the Landau level is half-full.  In typical situations,  we will not find that  electronic states  in the Landau level are entirely empty at positions outside the edge state or entirely full inside the edge state.  Rather, the Landau level will have a certain number  of electrons $N^e_j$ in localized states outside the edge state, and a certain number $N_j^h$ of unoccupied localized states (holes) inside the edge state.  The quantities $N^h_j$ and $N^e_j$ are constrained to be integers, as they represent the occupations of localized states.

In our analysis we will neglect the electrons and holes localized between edge states, and consider only those that are localized in the bulk of the sample, where the filling factor is $\nu_{in}$. This neglect is justified below, towards the end of this section. 
We shall also assume that the electrons that are localized in the $\nu_{in}$ bulk region are weakly conducting and compressible over long time scales, so that we can view them as forming a metallic region of a uniform electro-chemical potential, whose number of electrons is quantized to an integer $N_L$.\cite{Evans+93}  Experimental support for this picture was found in [\onlinecite{Ilani+04}]. Both of these assumptions are further elaborated on towards the end of this section.

Within this model, then, the interferometer has a single discrete degree of freedom, $N_L$, and several continuous degrees of freedom $A_j$, describing the area (relative to a reference area) occupied by each of the edges that are coupled to the leads (the subscript numbers the edge state). As is always the case in the quantum Hall effect, charge density on the edge translates to an area enclosed by the Landau level. The phase $\theta$ is directly related to the area $A_I$ enclosed by the interfering edge state, as delimited by the points in the constrictions where there is tunneling between the partially transmitted edge states.   (The subscript $I$ stands for ``interfering").  
Specifically, the relation is  given in Eq. (\ref{thetaAI}), above. Alternatively, we can consider $\theta$ as a measurable quantity (mod $2 \pi$), and use (\ref{thetaAI}),  to form a precise definition of $A_I$.

\subsection{Macroscopic energy function}

We will now formulate the way by which we will calculate (\ref{intphase}) and its dependence on $B$ and $V_G$. Since the phase $\theta$ depends only on what happens in the $\nu_{in}$ bulk region, we find it useful to define an energy functional $E(N_{\rm L}, A_I)$ as the total energy of the system when $N_{\rm L}$ and $ A_I$ are specified, and the energy is minimized with respect to all other variables, including the fluctuating areas $A_j$ of any fully transmitted edge states. (The electrochemical potential $\mu$ of the leads is here taken to be zero.).

Let us consider small variations of $B$ about a given initial value $B_0$, at a fixed value of the gate voltage $V_G$.  For small variations in $N_L , A_I$, we may then expand the energy $E(N_{\rm L}, A_I)$ to quadratic order, and write
\begin {eqnarray}
\label{Enphi}
E  &= &   \frac{K_{I}}{2} \,  (\delta n_I)^2  +   \frac{K_L}{2} \, (\delta n_{\rm L})^2  \nonumber \\
& &  +  \,\, K_{IL}\,  \delta n_I  \,  \delta n_{\rm L},
\end{eqnarray}
where  $\delta n_L$ is the deviation of the number of localized electrons from the value that would minimize the energy if there were no integer constraint on $N_L$,  and $\delta n_I$ is the deviation of the charge on the interfering edge, in units of the electron charge,        from the charge that would then  minimize the energy.
More precisely,
\begin{equation}
\delta n_L=N_L + \nu_{in}\phi-\bar q
\label{deltanL}
\end{equation}
where $\bar q$ is the effective positive  background charge, in units of $|e|$, resulting from ionized impurities in the donor layer and additional charges on the surfaces and on metallic gates, as well as any fixed charges in localized states outside the interference loop.  We assume that $\bar q$ depends monotonically 
on the gate voltage. 
Furthermore, for weak backscattering,
\begin{equation}
\delta n_I  =B(A_I-\bar A)/\Phi_0 \, =  n_I - \phi,
\label{deltanI}
\end{equation}
where $n_I$ is the charge enclosed by the interfering edge state, ignoring the charges of the localized electrons and holes. 

When the gate voltage $V_G$ is varied with $B$ remaining fixed, the background charge $\bar q$ and the area $\bar A$ will vary. Their variation depends on the coupling of the gate to the interferometer, and we characterize it by two parameters:
\be
\beta= (B/\Phi_0) d \bar{A}/d V_G \, , \,\,\,\, \gamma = d \bar {q} / d V_G \, .
\label{gatecouplings}
\ee
The parameter $\beta$ describes the extent to which a variation of the gate voltage affects the area of the interferometer $A_I$ (and indirectly $\phi$), while $\gamma$ describes the way the gate affects the background charge in the bulk of the interferometer (and indirectly $N_L$).

Note that the  energy function (\ref{Enphi}) leads to an interference phase that is unchanged when $\phi$ varies by one. This change in  $\phi$ can be completely compensated in the energy function by changing $N_{\rm L}$ by the integer amount $- \nu_{\rm in}$, while $n_I$ changes by one.  The fixed value of $\delta n_I$ means that the area $A_I$ has not changed, but the phase $\theta$ has changed by $2 \pi$.  Such a  phase change has no effect on the value of $e^{i \theta}$.

\subsection {Fractional quantized Hall states}
Our considerations for the integer case can be easily extended to fractional quantized Hall states of the form
\be
\nu_{in} = I + \frac{p}{2ps+1} \, , \, \, \, \nu_{out} = I + \frac{p-1} {2s(p-1) + 1} \, ,
\ee
where $p$ and $s$ are positive integers, and $I\geq 0$ is an integer. These are filling fractions in the
range $I \leq \nu < I+1/2$, and we assume that they are correctly described by the standard composite fermion picture. Moreover, we assume that the backscattered excitation is the elementary quasiparticle of the state $\nu_{in}$, with charge $e^*_{in} = 1/(2sp+1).$
We again use a quadratic  energy function of the form (\ref{Enphi}), but now we have to modify  (\ref{deltanL}) and (\ref{deltanI}) and use (\ref{fractheta}) instead of  (\ref{thetaAI}) to describe the relations between $\delta n_I, A_I,N_L$ and $\theta$.

Specifically, the phase $\theta$ accumulated by an interfering quasi-particle is
\be
\frac {\theta}{2\pi} = e^*_{in} BA_I/\Phi_0 - 2N_L se^*_{in}  \, .
\label{theta-frac}
\ee
The first term is the Aharonov-Bohm phase, scaled down by the charge of the interfering quasi-particle, and the second term is the anyonic phase accumulated when one composite fermion goes around another.\cite{Halperin84,Blok90,Stern08}  
The statistical  phase $\theta_a$, which  appeared in  Eq. (\ref{fractheta}), is thus given by $\theta_a=-4 \pi se^*_{in}$.

An increase of the magnetic flux by one flux quantum introduces, on average, $\nu_{in}/e^*_{in}$ quasi-particles, hence modifying (\ref{deltanL}) to be
\be
\label{nLfrac}
\delta n_L = e^*_{in} N_L + \phi \nu_{in} - \bar{q}
\ee
Here $N_L$ is the net number of quasiparticles minus quasiholes, of charge $ e^*_{\rm in}$, inside the interfering edge state.

The relation between the area enclosed by the interfering edge and the charge contained in the corresponding composite fermion Landau level -- the modified version of (\ref{deltanI}) -- is,
\be
\label{nIfrac}
\delta n_I =  \Delta \nu  \, B  (A_I - \bar{A}) / \Phi_0 \,,
\ee
where $\Delta \nu \equiv   \nu_{in} - \nu_{out}$. The normalizations of $\delta n_I$ and $\delta n_L$ have been chosen so that they are measured in units of the electron charge.

As before, in the limit of weak back scattering, the resistance oscillation will be proportional to Re $<e^{i \theta}>$. Note that formulas for the fractional case reduce to those of the integer case if one sets $s=0$.

\subsection{Comments on the energy function}

The previous subsection has defined the model we will use for analyzing the interference term (\ref{intphase}) and its dependence on $B$ and $V_G$. Before carrying out this calculation, we pause to make some comments on the model.

\subsubsection{An alternative parametrization of the energy function}

The macroscopic energy function $E$ may be alternatively  described by an equivalent capacitor network.   If we introduce electrochemical potentials $\mu_I = \partial E / \partial (\delta n_I)$, and
$\mu_L = \partial E / \partial (\delta n_L)$, then the quadratic part of $E$ may be rewritten as
\be
e^2E  \,  =  \, \frac{C_I}{2} \, \mu_I^2 + \frac{C_L}{2} \, \mu_L^2 +
\frac{C_{IL}}{2} \,( \mu_I - \mu_L)^2 \, ,
\ee
where
\be
K_{I} = e^2 \frac{C_{L} + C_{IL}}{D},  \, \, \, K_L= e^2 \frac{C_{I} + C_{IL}}{D} , \, \, \,
K_{IL} = e^2 \frac{C_{IL}}{D} ,
\nonumber
\ee
\be
D= (C_{L} + C_{IL})  (  C_{I} + C_{IL}) - C_{IL}^2 .
\ee
The coefficients $C_L$ and $C_I$ may be interpreted as  effective capacitances to ground for the respective conductors, while $C_{IL}$ plays the role of a cross-capacitance. The effective capacitances result from a combination of classical electrostatics and quantum mechanical energies.

An advantage of rewriting the energy in this form is that it may be easier to understand the dependencies of the capacitance coefficients on the parameters of the system.  For example, we would expect the coefficients $C_I$ and $C_{IL}$ to be proportional to the perimeter $L$ of the interferometer, if the structure of the edge is held fixed.  The capacitance $C_L$ should be proportional to the area $\bar{A} $ of the island, if the center region is covered by a top gate with a fixed set-back distance.  On the other hand, we would expect $C_L$ to vary as $L \log L$, if there is no top gate and the nearest conductors are gates along the edges of the sample.

In the situation where the edge state is connected to leads in equilibrium at zero voltage, the equilibrium value of $\mu_I$ will be zero. Then the ground state energy will be given by
$E = (C_L+ C_{IL})   \mu_L^2 / 2e^2$, and we have $e^2 \delta n_L= \mu_L (C_L + C_{IL})$ and
$e^2 \delta n_I = - \mu_L C_{IL}.$

\subsubsection{Further justification for the model}

The major simplification involved in our model is the reduction of the number of degrees of freedom in the problem. In principle, the interferometer has edge states that form one-dimensional compressible stripes and a set of localized states between these stripes that may be either empty or full. Our model reduces the problem to two degrees of freedom, $A_I$ and $N_L$.

We use one number, $N_L$, to describe the number of localized states in the bulk of the interferometer based on the assumption  that electrons or holes in a localized state are localized in a one-body approximation, but are not completely immobile.  At any finite temperature, they have a non-zero conductivity due to processes such as  multi-particle hopping,  and we assume that they can readjust their relative positions continuously.  Thus, on a laboratory time scale,  the interior of the island should behave like a metal:   the $N_L$ charges arrange themselves  to give a constant electrochemical potential in equilibrium, within each class of localized states.\cite{Evans+93}    As a result  of the integer constraints on the total occupation numbers, however, there  can be small differences between the electrochemical potentials of the localized states and that of the adjacent edge states or leads.

We neglect the degrees of freedom associated with localized states between edge states. We assume again that the width $w$ of the region of non-uniform electron density near the edge of the sample is small compared to the overall radius to the island.  The area available for localized electrons or holes in the Landau level that  is partially filled in the center of the interferometer should be approximately $\bar{A}$, while the areas available for localized electrons or holes in any other Landau levels should be  of order $Lw$, which is much smaller.  Then the number of localized electrons or holes in any of these regions will be relatively small, and the energy cost of adding or subtracting a particle from one of them should be relatively high.    Thus we may generally neglect  fluctuations in these quantities at reasonably low temperatures.  The fluctuations in $N_L$ that do occur will arise normally from  changes in the occupation of the innermost partially full Landau level.

If the magnetic field $B$ or the gate voltage $V_G$ is varied by a sufficiently large amount, we do expect to encounter discontinuous changes in the occupations of  localized states other than those $N_L$ of the inner-most partially full Landau level.  These jumps should lead to  jumps in the phase $\theta$, which would appear as ``glitches'' in the  interference pattern. The analysis of periodicities given above apply, strictly speaking, only in the intervals between glitches.  The frequency of occurrence of  glitches should roughly correspond to the addition of one electron or one flux quantum in area $Lw$, which would be rarer by a factor of $Lw/{\bar A}$ than the oscillation frequencies we are interested in.  Also, in many cases, the coupling between the interfering edge state $\theta$ and a particular occupation number $N^h_j$ or $N^e_j$ may be sufficiently small that any glitches associated with changes in that occupation number would be unobservable.

Finally, in replacing the full energy function by the macroscopic function $E$, we have minimized the energy with respect to all continuous variables $n_j$ other than that of the partially transmitted edge state, i.e., we have ignored the effects of thermal fluctuations in these variables.  This neglect is justified for the continuous variables, because they enter the energy in a quadratic form, so their thermal fluctuations add only a constant to the energy.

\section {Aharonov-Bohm and Coulomb-Dominated regimes in the weak backscattering limit\label{weak-calc}}

We now have Eq.~(\ref{intphase}) for the resistance in the weak backscattering limit in terms of the interference phase $\theta$. We also have  Eqs.~(\ref{thetaAI},\ref{fractheta}, \ref{theta-frac}) for $\theta$ in terms of the degrees of freedom $A_I,N_L$, and the energy function (\ref{Enphi}) for the energy and its dependence on $B$ and $V_G$. In this section we make use of these expressions to calculate several thermal averages. First, we calculate  the abrupt phase jump $2 \pi \Delta$ that occurs when the number of localized electrons (or quasi-particles) varies by $-1$. Then, we calculate the magnetic field and gate voltage dependencies of $<e^{i \theta}>$  at high temperatures, and show that in that limit the interferometer shows either AB or CD behavior, depending on the value of $\Delta$. Finally, we turn to the case where AB and CD behaviors mix together, and develop the tools needed to analyze this case, at low temperatures as well as high.

\subsection{Continuous and abrupt phase evolution}

As  the energy function is quadratic with respect to the continuous variable $A_I$, the average $A_I$  is
the one that minimizes the energy function. For a fixed  number
$N_L$ of localized charges, we obtain
%
\begin{equation}
{\theta \over 2 \pi} \ = \ e^*_{in} \phi - 2se^*_{in}N_L -{K_{IL} \over K_I} \frac{1}{e^*_{\rm out}}\left[ e^*_{in}N_L  + \nu_{\rm in} \phi - \overline {q} \right]  \ \ .
\label{averagephase}
\end{equation}

The abrupt phase jumps $2 \pi \Delta$  associated with a change of $N_L$ by $-1$ can be read out from (\ref{averagephase}). For an IQHE interferometer we find
\be
\Delta = \frac {K_{IL}}{K_I} = \frac {C_{IL}}{C_L + C_{IL}} \, .
\label{Delta}
\ee
When there is no bulk-edge coupling $K_{IL}=0$ the interference phase is unaffected by $N_L$. When the bulk-edge coupling is strong, the jumps are unobservable, since $\Delta=1$.

For an FQHE interferometer, we have
\be
\Delta = \frac {K_{IL}}{K_I} +2e^*_{in}s \left (1-\frac {K_{IL}}{K_I}\right )
\label{DeltaF}
\ee
Now, if $K_{IL}=0$,  then  $ 2 \pi \Delta$ is the phase jump associated with the fractional statistics of the quasi-particles. When the bulk and the edge are coupled, the phase jumps reflect both the change of the area $A_I$ caused by the introduction of quasi-particles and the fractional statistics.
In the limit of strong coupling, where $K_{IL}=K_I$, the phase jump becomes unobservable, just as in the integer case. Now, if there is a change of -1 in $N_L$, corresponding to the introduction of a quasi-hole in the bulk, the area $A_I$ will  increase by $(e^*_{in} / \Delta \nu) (\Phi_0/B)$.  This is the area necessary to accommodate the charge of the quasi-hole, and is also the area necessary for the accumulated phase to grow by $2\pi$.

\subsection{Magnetic field dependence}

Next, if the parameters entering (\ref{Enphi}) are known, we may calculate the thermal expectation value
\be
\label{mean}
<e^{i \theta}>  = Z^{-1} \sum_{N_L} \int_{- \infty}^{\infty} d A_I e^{-E/ T} e^{i \theta} \, ,
\ee
with the partition function $Z$ given by
\be
Z = \sum_{N_L} \int_{- \infty}^{\infty} d A_I e^{-E / T}       \, .
\ee
Since $E$ is a quadratic function of its variables, the integration over $A_I$ is trivial.  The sum over the discrete variable $N_{L}$ can be handled by using the Poisson summation formula and taking the  Fourier transform. Thus we may write
\be
\sum_{N_L= -\infty} ^ {\infty}
= \int_{- \infty} ^{\infty} d N_L \sum_{g=-\infty}^{\infty}  e^{-2 \pi i  N_L (g-1)}
\ee
Using this formula, one may perform the integrations over $N_L$ in the numerator and denominator of (\ref{mean}).
The formulas  simplify at high temperatures, where the partition function $Z$ becomes independent of $\phi$, and we may concentrate on the numerator of (\ref{mean}). We then find that
the expectation value can be written in the form
\be
<e^{i \theta}> \, =   \sum_{g = -\infty} ^{\infty}  D_m e^{ 2 \pi i \, m \phi} ,
\label{hightintphase}
\ee
where  $m(g)=-\frac{\nu_{out}}{e^*_{out}}+g\frac{\nu_{in}}{e^*_{in}}$, as in Eq. (\ref{allowedms}).

The coefficients $D_m$ may be written as
\be
D_m = (-1)^{g+1} |D_m| \exp \left[2\pi i \bar{q} \left( \frac {e^*_{in}-m}{ \nu_{in} } \right) \right] \, ,
\label{Dm}
\ee
with
\be
|D_m| =  e^{-2 \pi^2 T/ E_m}
\label{Dmabs}
\ee
and
\be
\frac{1}{E_m} = \frac {1}{(e^*_{out})^2  K_I }  +   \frac{(g - 1 +\Delta)^2 K_I}{(e^*_{in})^2(K_I K_L - K_{IL}^2)}
\label{oneoverem}
\ee
Remarkably, Eq. (\ref{oneoverem}) identifies the most dominant Fourier component of the resistance in the high temperature limit, and displays its relation to $\Delta$: in the integer case and for fractions where $m=1$ is allowed (i.e., for fractions in with $\nu_{in} < 1/2$), if $-1/2 < \Delta < 1/2$ the interference is dominated by the AB component, with $g=1, \, m=1$. In contrast, if $1/2 < \Delta < 3/2.$ it is dominated by the
 CD component,  with  $g=0$, $m=1 - (\nu_{in}/e^*_{in})$

We note that the plausible assumption of a positive cross capacitance $C_{IL}>0$ leads to the restriction $0< \Delta < 1$.  We will then find $\Delta < 1/2$ if and only if $C_{IL} < C_{L}$. We also remark that the energy $E_m$ for the CD term is related to the capacitances by $E_m = (e^*_{in})^2 / (C_L+C_I)$.  The denominator here may be thought of as an effective capacitance resulting from the electrostatic and quantum capacitances of the combined system of the localized charges and the interfering edge state, if the edge state is disconnected from the leads.

\subsection{Gate voltage dependence}

A variation of the gate voltage $V_G$ varies the phases of the Fourier components of $<e^{i \theta}>$
through its effect on $\bar q$ and $\phi$ in the phases in Eqs. (\ref{hightintphase}) and (\ref{Dm}).  There are two origins to this dependence - the effect of the gate voltage on the area of the interference loop $\bar A$ and its effect on the charge density in the bulk, and through it, on $N_L$. These two dependencies are described by the parameters $\beta,\gamma$ of (\ref{gatecouplings}).

For small variations $\delta V_G$ and $ \delta B,$  we see that
$D_m e^{2 \pi i m \phi} $ varies proportional to
\be
\exp \left( 2 \pi i \left[ \, \delta V_G  (  \alpha_m + \beta m ) +  m \delta B \frac { \bar{A}}{\Phi_0}
\right]  \right)  \, ,
\ee
where the term proportional to $\beta$ originates from the area change induced by the gate, and the term proportional to
\be
\alpha_m =  \gamma (e^*_{in} - m)/ \nu_{in} \, .
\ee
originates from the effect of the gate on the bulk background charge.

For the integer case, we see that lines of constant slope in the AB regime will have
$dV_G/dB = -\bar{A}/ \Phi_0 \beta$ , while in the CD regime, the lines of constant slope will have
$dV_G/dB = f_T \bar{A}/ \Phi_0 ( \gamma - f_T \beta)$.

We expect that  applying  positive voltage  to a side gate should tend to increase the area $\bar{A}$, so that the coefficient $\beta$ should be positive.  To estimate $\gamma$, let us first consider a model in which there is a constant electron density in the interior of the interferometer, except for a thin region around the edge, and let us imagine that the effect of $\delta V_G$ is to alter the location of the edge, without changing its density profile, and without changing the electron density away from the edge.  In this case we would find
$\gamma= \bar{\nu} \beta$,  where  $\bar{\nu}\geq (f_T+1/2)$ is the  filling factor in the interior.   In reality,  we would expect that positive $\delta V_G$ will increase the average density inside, so that $\gamma$ should be even larger.  Thus we expect that the slope of the constant phase lines will be negative for the  AB stripes but positive for the CD stripes.

\subsection{Low temperatures}

Although at high temperatures we need only consider one Fourier component, at lower temperatures, particularly if $\Delta$ is close to 1/2, the $g=0$ and $g=1$ components may both be important.  Then a color-scale map of the interference signal versus $B$ and $V_G$ will show lines of both slopes, with a resulting pattern of a checker-board type, as seen in Fig 2.  Even if both slopes are present, however, the eye will tend to pick out only the stronger component, if there is a big difference in the amplitudes, as in panels 2b and 2d. 

At still lower temperatures, higher harmonics with $g>2$ and $g<0$ will also appear.  In general one must take into account that $Z$ in the denominator of (\ref{mean}) depends on $\phi$.  Let us  expand $Z$   as
\be
Z = \sum_{g= - \infty}^{\infty} z_g e^{2 \pi i g ( \nu_{in} \phi - \bar{q})    / e^*_{in}} \, .
\ee
(The coefficients $z_g$ fall off exponentially with increasing temperature, except for $z_0$, which is simply proportional to $T$.)
The Fourier components of $<e^{i \theta}>$ will then be a convolution of the Fourier components of  $Z^{-1}$ with the Fourier coefficients  obtained from the numerator of (\ref{mean}), which are given by (\ref{Dm}) and (\ref{Dmabs}).
We see that this does  not introduce any new Fourier components into the function, but it can affect the relative weights of the different harmonics.

In the limit of low temperatures, the phase $\theta$ becomes a saw-tooth function of  $\phi$, for fixed $V_G$, and we can simply  evaluate the Fourier coefficients of $e^{i \theta}$. Up to a constant phase factor, we find that for the allowed values of $m$, the coefficients $D_m$ may still be written in the form
(\ref{Dm}), but now
\be
|D_m| =\frac {\sin(\pi \Delta)}{\pi (\Delta + g -1)}
\ee
We see  that the
CD component  $(g=0)$ will be largest  if $1/2 < \Delta \leq 1$, and the component  $(g=1)$ will be largest if $0\leq \Delta < 1/2$,  at T=0 as well as at high temperatures.

In our discussions of the temperature-dependence of the interference signal, we have taken into account only classical fluctuations, ignoring quantum fluctuations, which can be important on energy scales larger than $k_BT$.   In the FQHE case, quantum fluctuations lead to a renormalization of the tunneling amplitudes, which will typically cause the individual reflection amplitudes $r_1,r_2$ to decrease with increasing temperature,  as a power of $1/T$, in the weak backscattering regime.\cite{Wen91} At high temperatures, this  decrease should  be less important than the exponential decrease of the interference signal arising from classical fluctuations, predicted by Eq. (\ref{Dmabs}), but the power-law dependence should be taken into account at lower temperatures.  
If one defines a normalized interference signal by dividing the interference term by the total backscattered intensity, $\propto (|r_1|^2 + |r_2|^2)$, then the low temperature power-law dependence should be cancelled.\cite{Chamon+97}   Quantum fluctuations do not lead to a power law dependence of the normalized interference signal on length of the interferometer, in the limit of vanishing temperature and vanishing source-drain voltage.\cite{Chamon+97}

\subsection{Two-dimensional description} \label{2Dsec}

For a proper analysis of the regime where the CD and AB lines co-exist, we need to introduce a two-dimensional Fourier transform of $\delta R$ with respect to $B$ and $V_G$, rather than the Fourier transform with respect to $\phi$ at fixed $V_G$, which we  have employed so far. One finds that  the periodic pattern can be expanded in terms of a set of  ``reciprocal lattice vectors'' $\vec{G}_{gh} \equiv (G_{gh}^{(b)} , G_{gh}^{(v)})$, where $g$ and $h$ are integers,  with
\be
\label{Ggh}
\vec{G}_{gh} = g\vec{G}_{10} +  h \vec{G}_{01} \,
\ee
\begin{eqnarray}
\vec{G}_{10} & = &
2 \pi  \left(\frac{\nu_{in}}{e_{in}^{*}}\frac{\bar{A}}{\Phi_{0}},\,\,\frac{\beta \nu_{in}-\gamma}{e_{in}^{*}}\right)\label{G10}\\
\vec{G}_{01} & = &
2 \pi \left(-\frac{\nu_{out}}{e_{out}^{*}}\frac{\bar{A}}{\Phi_{0}},\,\,\frac{\gamma - \beta \nu_{out}}{e_{out}^{*}}\right)
\label{G01}
\end{eqnarray}
and
\be
\label{mnsum}
\delta R(B, V_G)= \sum_{gh} R_{gh} e^{  i  (G^{(b)}_{gh}\, \delta B + G_{gh}^{(v)}\, \delta V_G)} \, .
\ee
The reality of $\delta R $ requires that $R_{gh} = R^*_{-g,-h}$.

The set of reciprocal lattice vectors may be derived by first removing the restriction that $N_L$ is an integer. Regardless of the values of $K_I,K_L,K_{IL}$, the energy can then be minimized by choosing $A_I$ and $N_L$ so that $\delta n_I = \delta n_L = 0$. using (\ref{nIfrac}) and (\ref{nLfrac}).  If we then calculate  changes in $\theta$ using (\ref{theta-frac}), we find that
 $\delta \theta = G^{(b)}_{11} \delta B + G^{(v)}_{11} \delta V_G$, while
   $\delta N_L = -(G^{(b)}_{10} \delta B + G^{(v)}_{10} \delta V_G )/ 2 \pi$, with $\vec{G}_{gh}$ defined as in  (\ref{Ggh}) - (\ref{G01}).
Here,  we have used the relations  
$\Delta\nu=e^*_{ in}e^*_{ out}$ and $2s=(e^*_{out}-e^*_{in}) / \Delta \nu$.

In the limit of weak back scattering, the only reciprocal lattice vectors with non-zero amplitudes have $h=\pm 1.$  For $h=1$, the coefficients $R_{gh}$ may be related to the coefficients $D_m$ defined previously, with $R_{g,1} \propto r_1 r_2 D_{m}$, where  $m$ is related to $g$ by Eq. (\ref{allowedms}) and $r_1,\, r_2$ are the bare reflection amplitudes at the two constrictions.  For $h=-1$, the coefficients are the complex conjugates of $R_{-g,1}$.

When one goes beyond weak backscattering, as discussed below, one finds harmonics at reciprocal lattice vectors which are arbitrary sums of the ones present in the weak back scattering limit. Thus, one may obtain contributions at all integer values of $h$, including $h=0.$

Using the two-dimensional description, we may readily extend our  analysis to the situation where one cannot neglect the dependence of the secular area $\bar{A}$ on the magnetic field $B$. In this case, we should also take into account the change in the ``background charge'' $\bar{q}$ resulting from the change in $\bar{A}$.  We define two dimensionless parameters,
\be
\lambda = - \frac{B}{\bar{A}} \frac{\partial{\bar{A}}}{\partial B} \, , \,\,\,\,\,
\eta =  - \frac{\Phi_0}{\bar{A}} \frac{\partial{\bar{q}}}{\partial B} \, .
\ee
Then the formulas for the fundamental reciprocal lattice vectors should be replaced by
\begin{eqnarray}
\frac{\vec{G}_{10}}{2 \pi} & = &
\left[ \left(\frac{\nu_{in}(1-\lambda)+ \eta}{e_{in}^{*}} \right)\frac{\bar{A}}{\Phi_{0}},\,\,\frac{\beta \nu_{in}-\gamma}{e_{in}^{*}} \right]   \label{G10F}\\
\frac{\vec{G}_{01}}{2\pi} & = &
\left[ - \left(\frac{\nu_{out}(1-\lambda)+ \eta }{e_{out}^{*}}  \right)   \frac{\bar{A}}{\Phi_{0}}    ,\,\,\frac{\gamma            - \beta \nu_{out}}{e_{out}^{*}} \right] \label{G01F}
\end{eqnarray}
If the field $B$ is varied while the gate voltage $V_G$ is held fixed, the field periods associated with the AB term $(g,h)=(1,1)$ and the CD term  $(g,h)=(0,1)$ are given, respectively, by
\be
\bar {A} \, (\Delta B)_{\rm{AB}} = \frac {\Phi_0 } {(1-\lambda)\left(\frac{\nu_{in}}{ e^*_{in} }
- \frac {\nu_{out}}{ e^*_{out}} \right)   + \eta \left( \frac{1}{ e^*_{in}} - \frac{1}{ e^*_{out} } \right)   }\, ,
\ee
\be
 \bar {A} \, (\Delta B)_{\rm{CD}} = -\frac { e^*_{out}\Phi_0}{\eta+ \nu_{out}(1-\lambda)} \, .
\ee
If $\eta \neq 0$, the two periods will generally be incommensurate. Then when the magnetic field is varied at constant gate voltage, the resistance will not be a periodic function of $B$, but rather quasi-periodic.  To obtain a periodic variation, one must  vary $B$ and $V_G$ simultaneously, along a line of appropriate slope.

As a simple example, let us assume that $\bar{A}(B,V_G)$ is determined by a contour in the zero-field electron density $n(\vec{r})$ where $n \Phi_0 / B = (\nu_{in}+\nu_{out})/2$, and let us assume that $\bar{q}$ is equal to the integral of this density inside the area $\bar{A}$.  Then we find
\be
\eta = \frac {\lambda (  \nu_{in}+\nu_{out})}{2} \, ,
\ee
\be
\label{lambda}
\lambda=\frac { 1   }{ \bar{A} }  \oint \frac{ n(\vec{r}) dr}{|\nabla n|} \, ,
\ee
where the integral is around the perimeter of the area $\bar{A}$. We see that $\eta$ and $\lambda$ will vanish in the limit where the length scale for density variations at the edge is small compared to
 the radius of the island, (assuming that the density in the bulk is not too close to density  at the interfering edge state).

\section{Intermediate Backscattering\label{intermediatebackscattering}}

If one goes beyond the lowest order in the backscattering amplitudes $r_1$ and $r_2$, the above analysis must be modified in several respects. In this Section we confine ourselves to the IQHE case; we come back to the
FQHE in the next Section, for the regime of strong back-scattering.

The most obvious change from the weak backscattering limit  is that the interference contribution to the resistance $R_D$ is no longer simply proportional to $\rm{Re} [r_1 r_2^* e^{i \theta}]$.  To be specific, let us consider the case of symmetric constrictions, so that $r_1 = r_2$.   We may write
$R_D^{-1} = (f_T+1 - P_R) \, (e^2/h)  $, where $0<P_R<1$ is the probability that an incident electron in the partially transmitted  edge state will be  reflected by the interferometer region.
If we continue to define $\theta$ as the phase accumulation around the interferometer loop for an electron at the Fermi energy, then the full expression for  $P_R$ is
\be
P_R =  2  |r_1|^2 \frac {1 + \cos \theta}{1 + |r_1|^4 +2 |r_1|^2 \cos \theta}
\label{Rtheta}
\ee
If we expand this in powers of $r_1$, we find terms of order $|r_1|^4$ multiplying $\cos^2 \theta$, etc., which we may understand as contributions from
electrons that undergo multiple reflections and therefore traverse the circuit more than once.  Such terms will add additional harmonics of $e^{2 \pi i \phi}$ to the reflection coefficient, and in principle all harmonics will be present. However, the underlying period will not be affected.  Moreover, at least at high temperatures, the higher harmonics should fall off faster than the principal  AB component ($\propto e^{2 \pi i \phi}$) or the principal CD component ($\propto  e^{-2 \pi i f_T \phi}$) and should not be very noticeable.

In the presence of a significant reflection probability, one should also take into account the fact that in this case the number of electrons enclosed by the partially transmitted  edge  state is no longer precisely equal to $ \theta / 2 \pi $.  This  follows from the the Friedel sum rule,  which states that  $\rho(\epsilon)$, the density of states for the Landau level inside the interferometer at energy $\epsilon$ may be written as
\be
\rho({\epsilon}) =  \frac {1}{ \pi}  \frac { \partial ( \eta_+  + \eta_-)} {\partial \epsilon} \, ,
\ee
where  $\eta_{\pm}$ are the phase shifts, derived from  the eigenvalues $e^{2i \eta_{\pm}}$  of the $2\times 2$ $S$-matrix for transmission through the interferometer.   Explicitly, the eigenvalues are given by
\be
e^{2i \eta_{\pm}} = \frac { (1-|r_1|^2)e^{i \theta} \pm i |r_1| (e^{2i \theta} +1)}
{1 + |r_1|^2 e^{2 i \theta}} \, ,
\ee
and the phase shifts are required to be continuous functions of the energy $\epsilon.$
For $|r_1|^2  \neq 0$, this gives an oscillatory contribution to the phase shifts, and an oscillatory contribution to the density of states.
Since the electron number $n_I$ is the integral of $\rho(\epsilon)$ up to the Fermi energy, it will also acquire an oscillatory part. Specifically we may write
\be
n_I = \pi^{-1} ( \eta_+  + \eta_-) +{\rm{const}} = (2 \pi)^{-1} \theta + f(   \theta) \, ,
\label{nI}
\ee
where the phase shifts are evaluated at the Fermi energy, and $f(\theta)$ is  periodic,  with period $2 \pi$.

The oscillatory  contribution to $n_I$ will also be manifest when one varies the magnetic field, the gate voltage, or the electrochemical potential $\mu$.  For an interacting system, where the number of electrons $n_I$ associated with the interfering edge state is coupled to other variables, such as $N_L$,  or even to continuous variables such as the number of electrons in fully transmitted edge states, an oscillatory component of $n_I$ will lead to an additional oscillatory component to the energy $E$, which should be taken into account when evaluating the thermal average of 
$P_R$.  Again, we see that these effects can lead to additional oscillatory contributions at harmonics of
the basic periods, giving rise to  non-zero amplitudes at  arbitrary reciprocal lattice vectors $\vec{G}_{gh}$, but they should not change the fundamental frequencies  $\vec{G}_{10}$ and   $\vec{G}_{01}$.

We can treat the case of intermediate (or strong) back scattering within our general model if we make a few  modifications of the definitions.  We continue to use the energy formula (\ref{Enphi}), with the definitions (\ref{thetaAI}) and (\ref{deltanL}) for $\theta$ and $\delta n_L$.  We continue to define $\delta n_I \equiv n_I - \phi$, as in (\ref{deltanI}) but we can no longer equate this to $B(A_I-\bar{A})/\Phi_0$. Instead, we must compute $n_I$ using
(\ref{nI}).  Finally, we must calculate $< P_R>$  by averaging (\ref{Rtheta}) with the weight $e^{-E/T}$, integrating over $A_I$ and summing over $N_L$.

\section{Strong Backscattering\label{strongbackscattering}}

It is interesting to explore the behavior of the interferometer at low temperatures in the limit of strong backscattering, where the amplitude $r_1$ is close to unity.  For the case $r_1=r_2$, when $\theta$ is an odd integer  multiple of $ \pi$ a resonant tunneling occurs, and $P_R=0$.  Then, for non-interacting electrons, at  large $r_1$,  we would find that the the reflection probability $P_R$ is close to unity most of the time, but there would be a series of values of the magnetic field, or of the gate voltage, where in a narrow interval, $P_R$ drops to zero.   The actual vanishing of $P_R$ is special to the case where $r_1=r_2$, but even for an asymmetric case, one would find  reductions in $P_R$ in the vicinity of the points where $\theta$ is an odd multiple of $\pi$.

We now analyze the effect of interactions between electrons on these transmission resonances, and in particular on the flux spacing $\Delta\phi$ between transmission resonances. Interestingly, we find that this spacing is different for interferometers in the AB and CD regimes.

In the limit of strong backscattering the charge enclosed in the $\nu_{in}$ area is almost quantized in units of $e^*_{\rm out}$. The condition for  transmission resonance,  that $\theta$ is an odd multiple of $\pi$, is  also the condition for a degeneracy of the energy for two consecutive values of the charge on the interfering edge. We now formulate this condition in terms of our energy functional and explore the magnetic field spacings between such resonances.

We start with the integer quantum Hall regime. Let $N_o = n_I +N_L$ be the total number of electrons in the higher Landau levels enclosed by the (almost-closed) interfering edge channel, excluding electrons in the $f_T$ filled Landau levels that correspond to the totally transmitted channels.
The energy of the system is then
\begin{eqnarray}
E(N_o,N_L)=\frac{K_I}{2}(N_o-N_L-\phi)^2+\nonumber \\ K_{IL}(N_o-N_L-\phi)(N_L+(f_T+1)\phi-{\bar q})\nonumber \\ +\frac{K_L}{2}(N_L+(f_T+1)\phi-{\bar q})^2
\label{energy-int}
\end{eqnarray}
An increase of $\phi$ by one decreases $N_L$ by $(f_T+1)$ and increases $N_o$ by $f_T$.
Resonant transmission occurs when there is a vanishing  energy cost for adding one electron to the closed edge, that is, a vanishing energy cost for varying $N_o$ by one while keeping $N_L$ fixed. Degeneracy points  where  $N_L$ changes by $\pm 1$ while $N_o$ is fixed will generally not lead to resonances, even though $n_I$ changes by $\mp 1$ at such points.  Although the $\theta$ will technically pass through an odd multiple of $\pi$ in this process, one expects that these transitions will generally happen discontinuously, so there is no point at which the resonance could be observed. Points where $N_L$ and $N_o$ increase simultaneously do not involve a change in $n_I$ and do not lead to transmission resonances.

In the extreme AB limit, where $K_{IL}=0$,  there are degeneracy points where $E(N_o,N_L)=E(N_o+1,N_L+1)$ separated on the $\phi$ axis by spacings $\Delta\phi=1/(f_T+1)$. These points do not, however, lead to resonances, since they involve a change in $N_L$. Degeneracy points that do lead to resonances occur when $E(N_o,N_L)=E(N_o+1,N_L)$, and the spacings between those is $\Delta\phi=1$, the flux period that characterizes also the weak backscattering regime of the AB limit.

In the extreme CD regime, $K_I=K_{IL}$, and  stability requires $K_L>K_I$. Then  jumps of $N_L$ are separated from jumps of $N_o$.  In an interval where $\phi$  increases by 1, there will be ${f_T}$  resonant events where $N_o$ decreases by one, while $N_L$ is fixed, and $(f_T + 1)$ separate events where $N_L$ increases by one while $N_o$ is fixed.
The resonances are thus separated by $\Delta\phi=1/f_T$. Again, this is the flux period that characterized the CD regime in the weak backscattering limit.

The difference in $\Delta\phi$ between the AB and CD limits characterize also the fractional case. In this case the bulk accommodates $N_L$ quasi-particles of charge $e^*_{in}$, and the total charge in the $\nu_{in}$ region is quantized in units of $e^*_{\rm out}$.
The charge on the interfering edge is  given by
\begin{equation}
n_I=N_oe^*_{\rm out}-N_Le^*_{in} \, .
\label{areaclosedfrac}
\end{equation}
Then, the energy functional becomes,
\begin{eqnarray}
E(N_o,N_L)=\frac{K_I}{2}(e^*_{\rm out}N_o-{e^*_{in}}N_L- \Delta \nu  \, \phi)^2+\nonumber\\ K_{IL}(e^*_{\rm out}N_o-{e^*_{in}}N_L- \Delta \nu \, \phi)(e^*_{in}N_L+\nu_{\rm in}\phi-{\bar q})\nonumber\\ +\frac{K_L}{2}(e^*_{in}N_L+\nu_{\rm in}\phi-{\bar q})^2
\label{energy-frac}
\end{eqnarray}

In the Coulomb dominated limit,  ${K_{IL}}/{K_I}=1$,  and the number of  transmission resonances that occur while  $\phi$  changes by one is equal to   $\nu_{\rm out}/e^*_{\rm out}$ .  This leads to
 $\Delta\phi=e^*_{\rm out}/\nu_{\rm out}$. The leading component in the Fourier transform of $P_R(\phi)$ would then correspond to the $g=0$ component of (\ref{allowedms}), just as in the weak backscattering limit.

 In the extreme Aharonov-Bohm limit, where $K_{IL}=0$, the structure of transmission resonances is more complicated, due to the difference between the elementary charges  $e^*_{\rm in}$ and $e^*_{\rm out}$. Just as in the weak backscattering case for the FQHE at $K_{IL}=0$, there is no single dominant value of $g$.  In the case of weak backscattering, this occurs because $\Delta = 2se^*_{in} \neq 0$, according to Eq. (\ref{DeltaF}).  Here we note that $e^*_{out}-e^*_{in} = 2s e^*_{in} e^*_{out}$.

Over all, we see that in the limit of strong backscattering,  in the CD regime,  
the number of peaks  in the transmission probability as we increase $B$ by one flux quantum is the same number $\nu_{\rm out}/e^*_{\rm out}$ as we obtained in the weak backscattering regime, consistent with the prediction that the period of the CD oscillations would not change as we vary $r_1$.  The strong back scattering limit may also be understood as a Coulomb-Blockade effect:  maxima in the transmission probability occur at points where the system consisting of the localized states and the almost-totally-reflected edge state is about to change from one integer value to another.

Typically, the reflection coefficient $r_1$ should increase from near zero to near unity as one decreases the electron density in the constrictions through the range where the filling factor $\nu_c$ at the center of the constriction decreases from slightly  below $\nu_{in}$ to slightly above $\nu_{\rm out}$.  For an ideal constriction, the variation in  $r_1$  should be smooth and monotonic. In real constrictions, however, the variation may be more complicated, as the Fermi-level may pass through one or more resonances due to tunneling through localized states in the constriction.

Our discussion of the variation in $r_1$  should also apply if $\nu_c$ is varied by changing the magnetic field $B$ rather
than by changing a gate voltage at the constriction. Again, the field periods for  the AB and CD oscillations should remain fixed as long as $\nu_{\rm out}/e^*_{\rm out}$ does not change. However, the parameter $\Delta$ which governs the relative strengths of the AB and CD contributions could conceivably change as the other parameters are varied.

Under some circumstances, if there is a large region of intermediate electron density within a constriction, the number of localized states in the constriction may become so large that there is  a large density of states at low energies associated with rearrangements of electrons in these  states. Then, backscattering through the constriction could become incoherent, either because of inelastic scattering from the low energy modes, or because the path length for tunneling is changed randomly due to thermally excited rearrangements of the localized states. We assume that this does not happen in the system of interest.

\section{Model with multiple edge states\label{model}}

In order to better understand how the presence of multiple edge states may affect the parameters entering the energy function (\ref{Enphi}), we discuss here some simplified models which may illustrate the physics.

We consider the integer case, with $f_T$ fully transmitted edge states. We define $\delta n_i$ to be the charge fluctuation associated with a fluctuation in area of the $i$-th edge state, for $1\leq i \leq N$, where $N=f_T+1$, while $\delta n_i= \delta n_L$,   for $i=N+1$.  The partially reflected edge state has $i=N$, so $\delta n_I = \delta n_N$

We may now write the quadratic part of the energy in the form
\be
E = \sum_{ij} \frac {\ka_{ij}}{2} \delta n_i   \, \delta n_j ,
\ee
where the sums go from 1 to $N+1$. We assume that the coupling constants  $\ka_{ij}$ are known, and we wish to find  the values of the coupling constants  $K_L, K_I, K_{IL}$ which entered our earlier computations. We wish to specify the values of $\delta n_L$ and $\delta n_I$, and minimize the energy with respect to the other variables. This means that for $1\leq j \leq N-1$, we have
\be
\sum_i \ka_{ji}  \delta n_i =0 \, .
\ee
The resulting energy will be quadratic in $\delta n_I$ and $\delta n_L$, and the coefficients may be identified with $K_I, K_L$ and $ K_{IL}$.

We illustrate further with two examples. In our first model, we consider a situation where
\be
\ka_{ij} = U + \ka_1 \delta_{ij} \, ,
\ee
for $1 \leq i, j \leq N$, and
$$\ka_{ij} = \ka_L \, ,  \mbox { for } i=  j = N+1 \, ,
$$
$$
 \ka_{ij} = V \, ,  \mbox { if either } i \mbox{ or }  j = N+1 \mbox{ but } i\neq j .
 $$
In this model, the interaction between the edge states is entirely determined by the total edge charge $  \sum_{j \leq N}  n_j$, and the interaction with  $N_L $ involves only that charge.    After some straightforward algebra,  on obtains the results
\be
K_I = \ka_I + \tilde{U} \, , \,\,\, \, \,\,\,\, K_{IL} = \tilde{V} \, ,
\ee
where
\be
\tilde{U}  \equiv  U - \frac{ f_T U^2 }{ \ka_1 + f_T U} \, ,
\ee
and $\tilde{V} = V \tilde{U}/U$.
We see from these results that $K_{IL} / K_I= \tilde{V} / (\ka_1 + \tilde{U}) $. If $V \leq U$ and $f_T > 0$,  this ratio is necessarily less than 1/2 ,  so the model will be in the AB regime.   For $f_T=0$, the model leads to the CD  regime if and only if $\ka_1 +U < 2 V$.

The second model we consider is the opposite extreme, where edges are coupled only to their nearest neighbors.  We choose the diagonal coupling constants $\ka_{jj}$ as the previous model, while for off diagonal couplings we choose $\ka_{ij} = \ka_{12}$, if  $|i-j|=1$, and $\ka_{ij} = 0$ otherwise.
Now, we find that  couplng to the fully transmitted edges renormalizes the coefficient $K_I$ but has no effect on $K_L$ and $K_{IL}$, which remain equal to $\ka_L$ and $\ka_{12}$, respectively.  In the case $f_T=1$, we find
\be
K_I = \ka_1 - \ka_{12}^2/ \ka_{1} \, .
\ee

 The value of $K_I$ will be reduced further with increasing $f_T$, but the value remains finite  in the limit of large $f_T$, where one finds
 \be
 K_I \to \frac{\kappa_1}{2} + \frac{ ( \ka_1^2 - 4 \ka_{12}^2)^{1/2}}{2}
 \ee
 We see that in this model, $K_I$ is reduced by up to a factor of 2 as a
result of coupling to additional edges. Stability
of the model, in the limit of large $f_T$,  requires that $\ka_{12}/\ka_1 < 1/2$, and we see that
$K_{IL}/ K_I <1$. At the same time, if $2/5 <\ka_{12}/\ka_1 < 1/2$, the ratio
$K_{IL}/ K_I$ will be greater than 1/2, for sufficiently large $f_T$, so
the system may be pushed from AB  into the CD regime.  Of course, the CD regime
could be reached more easily if the model is modified so that
the coupling $\ka_{N,N+1}$ between the localized charge and the partially
reflected edge state is made larger than the other coupling energies, or
if the diagonal element $\ka_{NN}$ is made smaller than the coefficients
$\ka_{ii}$ for the fully transmitted edge states.

For a  uniform edge of length $L$, we expect that the constants $\ka_{ij}$  should be proportional to $1/L$, for $1 \leq i,j < N $. If we write
\be
\ka_{ij} = \frac {2 \pi \hbar}{e^2 L} v_{ij} \, ,
\ee
then coefficients $v_{ij}$ have the dimensions of  velocity.  If we can neglect the response of all other degrees of freedom in the system, then fluctuations in the densities $\delta n_i$ for the edge modes will propagate with velocities that are given by the eigenvalues of the velocity matrix $v_{ij}$.  In actuality, however, the situation is more complicated, as coupling to charges in localized states may reduce the coupling constants $\kappa_{ij}$ and the propagation velocities by different amounts. The propagation velocities are only affected by rearrangements of charge or polarization that can take place on  time scale faster than the time $L/ v$ for a pulse to propagate around the interferometer, whereas the constants $\kappa_{ij}$ entering our formulas are defined for fluctuations on a longer time scale.

\section{Connection to previous theoretical and experimental works\label{Literature}}

Oscillations in the transport properties of quantum Hall devices, associated with interference effects,  were already observed in the 1980's, in both IQHE and FQHE regimes.\cite{Simmons89} 
  The possible importance  of Coulomb blockade effects\cite{Lee90} in these experiments, and of fractional statistics\cite{Kivelson90} for the FQHE situation, was noted by theorists around that time. Interpretation of the early experiments was difficult, however, as the interfering paths were not the result of a deliberate construction but were, presumably, the  result of random fluctuations in the doping density, whose geometry was not known.  In a typical case, one might see oscillations in the resistance of a micron-scale Hall bar, on the high-field side of a quantized Hall plateau, which might be attributed to backscattering through a ``dot" or an  ``anti-dot" inclusion, where the electron density was higher or lower than in the surrounding electron gas. The strength of tunneling into and out of the dot or anti-dot was generally assumed to be weak, and the oscillations were associated with resonances as additional electrons or quasiparticles were added to the inclusion.\cite{Jain-Kivelson}   In later years, improved experiments were carried out using  fabricated anti-dots with controlled areas, in which one could investigate systematically the dependence on magnetic field and on electron density, controlled by a back gate.\cite{Goldman-antidot}

Quantum Hall interferometers with the Fabry-Perot geometry  studied in the present paper  have been explored  
experimentally by several groups. In several  early works , 
Coulomb blockade effects in a dot {\em weakly}
coupled to leads were studied, in a region with several filled LLs. \cite{McEuen+92,Marcus+94}.
The crossover between AB and CD regime for a weakly coupled dot was analyzed in
[\onlinecite{Beenakker+91}]. Both integer and fractional quantum Hall interferometers in the absence of charging effects were discussed in [\onlinecite{Chamon+97}].

In an earlier experiment \cite{vanWees+89}, a strong dependence of the magnetic field period $\Delta B$  on the constriction filling factor was found but interpreted in terms of a magnetic field dependent interferometer area. In a reanalysis  \cite{CaZhGo05,Dharma-wardana+92} of that experiment, it was pointed out that under the assumption of a magnetic-field-independent interferometer area, the data agree with $\Delta B \sim 1/\nu_{in}$.
 
More recently, several groups have conducted systematic investigations of  interferometers of different sizes, with and without top gates, in which they could set the filling factor in the constriction independently of the density in the bulk, and data has been collected as a continuous function of both magnetic field and side-gate voltage.\cite{Zhang+09,Ofek+09}  
The AB regime, the CD regime and the intermediate regime were all observed in these experiments.
In the he CD regime, 
when  lines of equal $R_D$ were plotted in the $B-V_G$ plane, they were found to have positive slope  for $\nu_{out}  \neq  0$,  or zero slope for  $\nu_{out}  =  0$.  The CD flux period, in the integer case, was found to be   $\Delta\phi=1/\nu_{\rm out}$, independent of the strength of the backscattering. The AB regime, observed in the IQHE, was characterized by lines of equal $R_D$ that had negative slope in the $B-V_G$ plane  and flux periodicity of $\Delta\phi=1$. Intermediate regimes, where AB and CD behaviors combined together were also observed, giving a checkerboard pattern of the type seen in Fig. [2c].

In the fractional case flux periodicity of  $\Delta\phi=1/\nu_{\rm out}$ was observed in the cases $(\nu_{in},\nu_{\rm out})=(\frac{1}{3},0)$ and $(\frac{4}{3},1)$, where $\nu_{\rm out}$ is an integer and $\nu_{\rm in}$ a fraction. In the case of $(\frac{2}{5},\frac{1}{3})$, a period of 
$\Delta\phi={e^*_{\rm out}}/{\nu_{\rm out}}=1 $ was observed.\cite{Ofek+09}

In an earlier work\cite{RoHa07},  in which two of us analyzed the interference patterns in a Fabry-Perot interferometer,   a parameter $\Delta_x/\Delta$  characterizing the strength of bulk-edge coupling was introduced. In the present notation, it corresponds to the ratio
$K_{IL}/K_I$. Here, we have gone  beyond the approach of [\onlinecite{RoHa07}] by  studying the directions of
lines of constant phase and the temperature dependence of the interference terms, and by  allowing  for arbitrary strength
of backscattering.

A first principles approach to the study of interferometers was described in [\onlinecite{Ihna+08}]. Possibly due to
the approximations chosen in that approach, an influence of Coulomb interactions on the magnetic field
period of resistance oscillations was not found.

A situation in which  the area of the interfering loop is small
compared to the lithographic area, and where it is  highly dependent on the
magnetic field was discussed in  [\onlinecite{Siddiki10}].
 In this paraemter regime, the coefficient $\lambda$, defined in our Eq.~({\ref{lambda}), can be larger than  2, so  $(1-\lambda)^{-1}$ can be negative, with a magnitude smaller than 1.  This would cause the Aharonov-Bohm  constant-phase lines to have reversed slope, and a period smaller than one flux quantum.
However,
under this assumption, the magnetic field period would vary continuously, rather than being quantized at a flux
quantum divided by an integer, so this mechanism does not seem to explain the experimental findings
\cite{Zhang+09,Ofek+09}. Also, this would not explain the simultaneous appearance  of AB and CD lines, as observed in several cases.

The influence of anyonic statistics on magnetic field periodicities of Fabry-Perot interferometers was
discussed in [\onlinecite{Goldman07}], although the results obtained there disagree in some cases with
our findings.

Observations of a magnetic field superperiod,  corresponding to an addition of five flux quanta to the interferometer area have been reported in [\onlinecite{Camino+05,Camino+07a,Ping+09b}] for a sample in which the bulk is in a quantized Hall state with $\nu=2/5$, while the constrictions have filling fraction $1/3$ .  We do not have an explanation for these results. However, we do not accept the theoretical explanation put forth  in these papers or in  [\onlinecite{Goldman07}].  Although we agree with the arguments which show that addition of five flux quanta should leave the interference pattern unchanged, we believe this should also hold for the addition of a single flux quantum, in the physical model presented in these papers.

\section {Discussion and Conclusions\label{summary}}

In this paper, we have presented a general framework for discussing the electronic transport in a quantum Hall Fabry-Perot interferometer.
Our aim was to understand the oscillatory  dependence of the interferometer resistance $R_D$ on the magnetic field $B$ and voltage applied to a side gate $V_G$, when these parameters are varied by an amount large enough to change the number of flux quanta or the number of electrons by a finite amount, but small enough so that there is not a large fractional change in either the flux or the electron number.
A central assumption was that the resistance arises from the partial reflection of  one quantum-Hall edge state in the two constrictions. We also restricted our analysis to the  integer quantum Hall states or a subset of fractional states, where all modes at a given edge propagate in the same direction. 
Our understanding of the physics of the problem was described in general terms in the Introduction, Section~{\ref{Introduction},  and in detail in the body of the paper. In this summary we focus on the results we obtained.

We found that $\delta R$, the oscillatory part of $R_D$, is, in general, a two-dimensional periodic function in the plane of $B$ and $V_G$.  It is useful to describe this function in terms of its two-dimensional Fourier transform, which  means that we should specify a set of  reciprocal lattice vectors $\vec{G}_{gh}$ and  the associated amplitudes $R_{gh}$,
where $g,h$ are arbitrary integers,  and  $\vec{G}_{gh} = g\vec{G}_{10} +  h \vec{G}_{01}$.
Explicit formulas for the reciprocal lattice basis vectors $\vec{G}_{10}$ and $\vec{G}_{01}$ were given in Subsection \ref{2Dsec}, in terms of the smoothly varying secular area $\bar{A}$ enclosed by the interfering edge mode, the filling factors $\nu_{in}$ and $\nu_{out}$ of the quantum Hall states separated by this edge mode,  and parameters $\beta, \lambda, \gamma, \eta$ describing the derivatives of $\bar {A}$ and the enclosed ``background charge'' $\bar{q}$ with respect to $V_G$ and $B$. Our most general expression for $\delta R$ is 
\be
\label{mnsum}
\delta R(B, V_G)= \sum_{gh} R_{gh} e^{  i  (G^{(b)}_{gh}\, \delta B + G_{gh}^{(v)}\, \delta V_G)} \, ,
\ee
with 
basis vectors given by (\ref{G10F}) and (\ref{G01F}).
However, in  cases where the radius of the interferometer is large compared  to the widths of the density transition regions at the edges, one may be able to neglect the magnetic-field dependence of $\bar{A}$ and $\bar{q}$, in which case $\lambda$ and $\eta$ may be set equal to zero. Then the  basis vectors $\vec{G}_{10}$ and $\vec{G}_{01}$ are given by the simpler expressions (\ref{G10}) and (\ref{G01}).

For the remainder of this summary, we shall  limit ourselves to the case $\lambda=\eta=0$. Then,  if $V_G$ is held fixed, we find that $\delta R$ is a periodic function of the magnetic field, with a fundamental period corresponding to the addition of one flux quantum to the area $\bar {A}$. However, the fundamental period may not have the largest Fourier amplitude, so the most visible oscillations may correspond to a harmonic, with a period that is the fundamental period divided by an integer.

For non-interacting electrons, in the integral quantized Hall effect, the observed interference pattern will reflect the fundamental  Aharonov-Bohm period, where the phase increases by $2 \pi$ when the dimensionless magnetic flux  $\phi \equiv  B \bar{A}/ \phi_0$ changes by one, due to variation of $B$ or of $V_G$ or both.  In our current notation, this means that  non-zero Fourier components $R_{gh}$ will  correspond to reciprocal lattice vectors where $g=h$.  In the case of weak backscattering at the constrictions,  or at high temperatures, the oscillations are simply sinusoidal, and the dominant contributions are   $R_{11}$ and its conjugate $R_{-1,-1}$.
On a a color-scale map of $\delta R$ in the $B-V_G$ plane, AB oscillations would appear as a series of parallel stripes with negative slope. For stronger backscattering, at low temperatures, we may get higher Fourier components due to multiple scattering events across the two constrictions.

In the case of  weak backscattering, Fourier components at additional reciprocal lattice vectors can arise from electron-electron interactions.  For the integer QHE, this is due to the Coulomb interaction between electrons on the interfering  edge state and localized electrons or holes which exist in the bulk of the interferometer. Because the  number of localized particles  is required to be an integer, the net number of localized particles $N_L$ will jump periodically, as $B$ or $V_G$ is varied. Interactions with the edge then 
cause small variations $\delta A_I$ in the area $A_I$ enclosed by the interfering edge state, which will cause the actual number of flux quanta enclosed by  $A_I$ to fluctuate about the nominal value $\phi$, and thus lead to an additional  modulation of the interference phase.  For FQHE states, there is an additional jump $\theta_a$  in the interference phase, arising from the fractional statistics,  whenever there  is a change in the number of localized quasiparticles.

If  the  Coulomb coupling between $N_L$ and the edge-state charge  is sufficiently strong, one finds that the dominant terms  in the Fourier expansion have $g=0$, both for the IQHE and the FQHE.  In this  limit, the color-scale map will show a series of stripes which have positive slope and $\Delta\phi=e^*_{\rm out}/\nu_{\rm out}$,  except when $ \nu_{out} = 0$, in which case the stripes will be horizontal on a $B-V_G$ plot.

In between the AB and CD limits, the two sets of stripes occur simultaneously, and a map of $\delta R $ will show a checkerboard pattern, as seen in Fig. 2, above.  The absolute strengths of the various Fourier coefficients will depend on the backscattering amplitudes in the individual constrictions and on the temperature, as well as on  three energy parameters, which we denote $K_I, K_L, K_{ IL}$.  
At high temperatures,  the  Fourier amplitudes will fall off exponentially  with $T$ with varying rates,   so generally a single pair of Fourier amplitudes will dominate at large $T$.  This may be either the AB term 
$(g=h= \pm 1)$ or the CD term $(g=0, h=\pm 1)$.  
Then, for a fixed gate voltage, the interference pattern in 
$\delta R$ will be a simple sine function of magnetic field, with  either the AB or CD period.

At lower temperatures, where many Fourier components maybe present the situation is more complicated.  We discuss here  the limit of weak backscattering, where $h$ is limited to $h=\pm 1$. 
Then, at low temperatures, one  finds that the phase $\theta$ of the interference path is a saw- toothed function of the magnetic field, varying linearly with $B$ most of the time, but with periodic  jumps by an amount $2 \pi \Delta$, which occur each time the number of localized quasiparticles $N_L$ changes by $-1$. There will be ${\nu_{in}}/{e_{in}^{*}}$  equally-spaced phase 
jumps per flux quanta change in the loop, which give rise to  Fourier components of $<e^{ i \theta}>$ with arbitrary values of $g$. For $h=1$, at $T=0$, the Fourier amplitudes vary with $g$   as 
$(g+\Delta-1)^{-1}$, according to Eq. (\ref{Dm}).  At higher 
temperature the jumps will be smeared, giving a more gradual change of $<e^{i \theta}>$
as $B$ is varied. This smearing causes  the Fourier amplitudes at the 
higher $g$'s to vanish exponentially.

We see from the above that  at low temperatures, the  AB Fourier amplitude will be larger than 
the CD amplitude if  the jump parameter $\Delta$ satisfies $0<\Delta<1/2$, while the reverse is true if $1/2 < \Delta < 1$. We find similarly at  high temperatures that  the AB component will be larger than the CD component if, and only if, $\Delta < 1/2$. 
For the IQHE $\Delta$ is purely a consequence of bulk-edge coupling, and it is equal to $K_{IL}/K_I$.  It vanishes in the extreme AB  iimit ($K_{IL} \to 0$) and approaches $1$ in the extreme CD limit ($K_{IL} \to K_I$). 
For the FQHE, the value of $\Delta$ depends on the statistical phase angle $\theta_a$
and the ratio $K_{IL}/K_I$, according to Eq. (\ref{DeltaF}), going from $\abs{\theta_a}/ 2 \pi $, in the absence of bulk-edge coupling, to $1$, when this coupling is strong. This suggests  the possibility that one could obtain a direct measure of $\theta_a$ by observing the discrete jump in the interference phase $\theta_a$  as an additional quasihole enters the interferometer at low temperatures.  In order to extract the value of  $\theta_a$, however,  one would have to independently find a measure of $K_{IL}/K_I$, or be able to vary $K_{IL}$ (say by varying the area of the interferometer) and extrapolate to $K_{IL} = 0$.

We have said little about the actual values of the
parameters  $\beta$ and $\gamma$ which determine the gate-voltage periods
of the AB and CD stripes, nor have we estimated the energy parameters
$K_I, K_L, K_{IL}$, which  determine the ratio between the AB and CD
amplitudes and the temperature dependence of these amplitudes.

One might try to estimate $\beta$ and $\gamma$ using a simplified model,
where the electron density in the sample depends on  $V_G$ but is
independent of $B$.  According to Eq. (11), this means that
if one considers a sample with fixed gate voltage, at various values of B,
corresponding to different bulk filling factors, the parameter $\beta$
will be proportional to $B$, while $\gamma$ will be independent of $B$.
Using Eq. (\ref{G01}) we find that the $V_G$ period for a CD stripe should be
equal to $e^*_{out} / (\gamma - \beta \nu_{out})$.  The filling factor
$\nu_{out} $ will depend on the magnetic field, but also may be varied  by
changing the voltage on the gates  defining the quantum point contact
constrictions of the interferometer.  It appears that the dependence of
the gate period on $B$ and $V_G$ predicted by these considerations is only
partly in agreement with experiment, and that significant effects are
omitted from this simple model.\cite{Zhang+09,Ofek+09}

Although the energy $K_L$ may be  largely determined by  the geometric
capacitance of the island, the parameters $K_{IL}$ and $K_I$ should be
sensitive to the detailed structure of the edge and are difficult to
estimate without  a detailed microscopic model and a numerical
calculation.  The values of these parameters should depend also on the
value of the magnetic field and on the setting of the constrictions, which
determines which edge mode is the interfering one.  For a dot of
sufficiently large area, covered by a top gate,  the parameter $K_{IL}$
should decrease inversely as the area,  so for an integer quantized Hall
state, one would be necessarily in the AB regime.  However, the converse
is not true;  for a small area  dot one could be in the CD or AB regime
depending on  details.   Further investigation of these points will be
left for future work.

Acknowledgments:  We acknowledge support from NSF grant DMR-0906475, from the Microsoft Corporation,  the BSF, the Minerva foundation, and the BMBF.  We have benefited from helpful discussions with  C.M.~Marcus, D.~McClure, Y.~Zhang, A. Kou,  M.~Heiblum, N.~Ofek, A. Bid, V.~Goldman, and R.~Willett.

\end{document}